\documentclass[]{article}

\usepackage[affil-it]{authblk}
\usepackage{a4wide}
\usepackage{color}
\usepackage[normalem]{ulem}
\usepackage[ruled,vlined]{algorithm2e}
\usepackage{hyperref}
\usepackage{amsmath}
\usepackage{amssymb}
\usepackage{graphicx}
\usepackage{amsfonts}
\usepackage{tabulary,booktabs}
\usepackage{multirow}
\usepackage{siunitx}
\usepackage[english]{babel}
\usepackage[sort&compress,square,comma,authoryear]{natbib}

\providecommand{\keywords}[1]{\textbf{Keywords ---} #1}

\definecolor{Red}{RGB}{212,28,48}
\definecolor{Green}{RGB}{81,153,74}
\definecolor{Blue}{RGB}{0,128,255}
\definecolor{Yellow}{RGB}{255,204,0}

\providecommand{\fp}{f_\mathrm{p}}
\providecommand{\ff}{f_\mathrm{0}}

\begin{document}
\graphicspath{ {./}{figures/} }
\bibliographystyle{plainnat}

\title{A Novel Greedy Approach To Harmonic Summing Using GPUs}

\author[1]{Karel~Ad{\'a}mek}
\author[2]{Jayanta~Roy}
\author[3]{Wesley~Armour \thanks{E-mail address: \texttt{wes.armour@oerc.ox.ac.uk}} }
\affil[1]{Faculty of Information Technology, Czech Technical University, Th\'{a}kurova 9, 160 00, Prague, Czech Republic.}
\affil[2]{National Centre for Radio Astrophysics, Tata Institute of Fundamental Research, Pune 411 007, India}
\affil[3]{Oxford e-Research Centre, Department of Engineering Sciences, University of Oxford, 7 Keble road, OX1 3QG, Oxford, United Kingdom}

\maketitle

\begin{abstract}
Incoherent harmonic summing is a technique which is used to improve the sensitivity of Fourier domain search methods. A one dimensional harmonic sum is used in time-domain radio astronomy as part of the Fourier domain periodicity search, a type of search used to detect isolated single pulsars. The main problem faced when implementing the harmonic sum on many-core architectures, like GPUs, is the very unfavourable memory access pattern of the harmonic sum algorithm. The memory access pattern gets worse as the dimensionality of the harmonic sum increases. Here we present a set of algorithms for calculating the harmonic sum that are suited to many-core architectures such as GPUs. We present an evaluation of the sensitivity of these different approaches, and their performance. This work forms part of the AstroAccelerate project \citep{AstroAccelerate} which is a GPU accelerated software package for processing time-domain radio astronomy data.
\end{abstract}


\section{Introduction}
\label{sec:introduction}
A convenient and computationally efficient way of detecting pulsars in time-domain radio astronomy is to use techniques based on the Fourier transform. The Fourier transform utilises the periodic nature of the pulsar's signal, but the use of the Fourier transformation is not without pitfalls. For example, variations in the pulsar's, otherwise periodic, emissions lead to degradation of the signal in the Fourier domain. This can occur when a pulsar is present in a binary system with another compact object (e.g. white dwarf, neutron star, black hole). In such a case, the Doppler effect changes the pulsar's period over the observing span resulting in drifts in the Fourier plane, and the use of an unaltered Fourier transformation does not aid in detection of the pulsar, and other methods must be used, such as FDAS \citet{Ransom2002:FDAS, Sofia:2018:FDAS}. 

The pulsar's signal can be described as a pulse train, where pulsar emissions are visible only in a portion of its period. The proportion between the part of the period when we see the pulsar's signal and the period duration is called the \textit{duty cycle}. For such signals, the Fourier transform distributes the power contained in the pulsar's signal into the fundamental frequency bin and multiple higher harmonic bins. The source of these harmonics is the combined effects of sampling and aliasing introduce by the discrete Fourier transform. In effect, some of the power contained in the pulse train is also added to the higher harmonics, which can be used to boost the pulsar's signal in the Fourier domain and increase the probability of detection. Moreover, this spread of power in higher harmonics increases for narrow duty cycle pulsars.

The incoherent harmonic sum (HS) aims to increase the pulsar's signal-to-noise ratio (SNR) by accounting power (either fully or partially) from increasing number of harmonics. Thus, the harmonic sum adds the signal present in the data and averages out the noise, increasing the pulsar's SNR proportional to $\sqrt{N_\mathrm{harmonic}}$ and consequently the probability of detection.

\citet{1969Natur.221..816T} first used the harmonic sum in 1969, only a few years after the discovery of pulsars by Antony Hewish and Jocelyn Bell Burnell  \citet{1968Natur.217..709H}. More recently, \citet{Ransom2002:FDAS} described how the duty cycle of the pulsar affects the number of significant harmonics. In the same work, \citet{Ransom2002:FDAS} also compared data from 600 pulsars and showed that a pulsar with a duty cycle below 0.1 could have more than ten significant harmonics. This shows the importance of the harmonic sum algorithm and it's role in pulsar detection. Furthermore, \citet{Yu_2018:HarmonicSum} performed a theoretical probabilistic analysis of the harmonic sum technique and proved the usefulness of this technique.

The one-dimensional harmonic sum algorithm is a standard in many software packages which process time-domain radio astronomy data, such as SIGPROC (\citet{Sigproc}) or the PRESTO software package (\citet{Presto}). However, none of these existing packages have the harmonic sum in a computationally accelerated form. The higher-dimensional harmonic sum is used in the Fourier domain acceleration search technique (FDAS) \citet{Ransom2002:FDAS}, or JERK search \citet{Andersen2018:JERK}, which searches for accelerated pulsars. 

In this work, we present a new algorithm to perform the harmonic sum and GPU implementations of this algorithm, evaluating its sensitivity and performance. Like the existing CPU implementations this algorithm is also approximate and does not perform an exhaustive harmonic sum. For example, the Lyne-Ashworth algorithm (described in \citet{Lorimer_Kramer:2004:handbook}) used in the SIGPROC package, widely used in pulsar detection, sums only harmonics which are powers-of-two. The harmonic sum algorithm used in the PRESTO software package \citet{Presto} also uses a subset of the harmonics. To compare our proposed algorithm, we have implemented a PRESTO like harmonic sum on the GPU and use this for our comparison. Algorithms are compared both in sensitivity and performance.  

We have integrated algorithms discussed in this work into AstroAccelerate\footnote{\url{https://github.com/AstroAccelerateOrg/}} \citet{AstroAccelerate}, a GPU accelerated real-time data processing software package for time-domain radio astronomy. The AstroAccelerate package contains GPU accelerated versions of commonly used algorithms for time-domain radio astronomy. Those are de-dispersion transform \citet{Armour2012:DDTR}, single pulse detection \citet{adamek2020:SPDT}, Fourier domain acceleration search \citet{Sofia:2018:FDAS} and accelerated JERK search \citet{Adamek2020:JERK}. AstroAccelerate is actively used as part of scientific pipelines like MeerTRAP \citet{MeerTRAP2021_a, MeerTRAP2021_b}, Greenburst \citet{GREENBURST2020}, GHRSS survey\footnote{http://www.ncra.tifr.res.in/$\sim$bhaswati/GHRSS.html} and it is being considered for the Petabyte FRB Search Project\footnote{\url{https://thepetabyteproject.github.io/}}.

This paper is organised as follows: in section~\ref{sec:hrms} we describe the harmonic sum, it's purpose and problems cause by sampling; in section~\ref{sec:algorithms} we introduce our new greedy harmonic sum algorithm and also describe existing algorithms; in section~\ref{sec:results} we present signal loss, recovered SNR, and compute performance of selected harmonic sum algorithms; in section~\ref{sec:discussion} we discuss achieved results, and lastly we conclude the paper in section~\ref{sec:conclusions}.


\section{Harmonic sum}
\label{sec:hrms}

The incoherent harmonic sum is given by equation
\begin{equation}
\label{eqa:generalHRMS}
X(f)_\mathrm{H} = \sum_{h=1}^{H}x\left(h\fp\right)\,,
\end{equation}
where $H$ is the total number of harmonics summed, $f$ is frequency, and $x$ is the continuous power spectra. The integer multiple $h$ will be called \textit{harmonic order} throughout this work. The harmonic sum algorithm produces a series of partial sums. For each sum it calculates its SNR and then selects the maximum. That is
\begin{equation}
    \label{eqa:SNR}
    \mathrm{SNR}(X(f)_\mathrm{H}) = \max_H\left(\frac{X(f)_\mathrm{H} - \mu_\mathrm{H}}{\sigma_\mathrm{H}}\right)\,,
\end{equation}
where $\mu_\mathrm{H}$ and $\sigma_\mathrm{H}$ are the mean and standard deviation respectively for partial sum of length $H$. The number of harmonics we need to sum is governed by the duty cycle of the pulsar we are looking for. 

When searching for a pulsar, where the frequency of the pulsar is unknown, the harmonic sum algorithm needs to perform partial sums up to the highest harmonic for every frequency bin (up to a limit, for example, dictated by observation or a theoretical model). 

The harmonic sum algorithm is performed after Fast Fourier Transformation (FFT) and further preprocessing, which includes spectrum whitening or power spectrum calculation which may include inter-binning as introduced by \citet{Ransom2002:FDAS}.

\subsection{Discrete harmonic sum}
When we have continuous frequencies, it is easy to pick the correct frequency of the higher harmonics as it is just an integer multiple of the pulsar's frequency $f_h=h\fp$. However, when dealing with discrete data choosing the correct frequency for summation is more complicated, and it is at the core of any harmonic sum algorithm. 

In the discrete data, the true frequency of the pulsar $\fp$ is represented by a \textit{fundamental frequency} $\ff$ which could be different from the pulsar's true frequency due to the finite resolution of the frequency representation of the observed data. Each frequency represented in the discrete Fourier domain has a \textit{frequency bin}, an integer index of that frequency. Therefore a bin that corresponds to the fundamental frequency is called the \textit{fundamental frequency bin}, and frequency bins that represent higher harmonics will be called \textit{harmonic bins}. We will use the terms bin and frequency interchangeably since there is a direct link between them.

\begin{figure}[htp]
    \centering
	\begin{minipage}[t]{.45\textwidth}
		\centering
		\includegraphics[width=\linewidth]{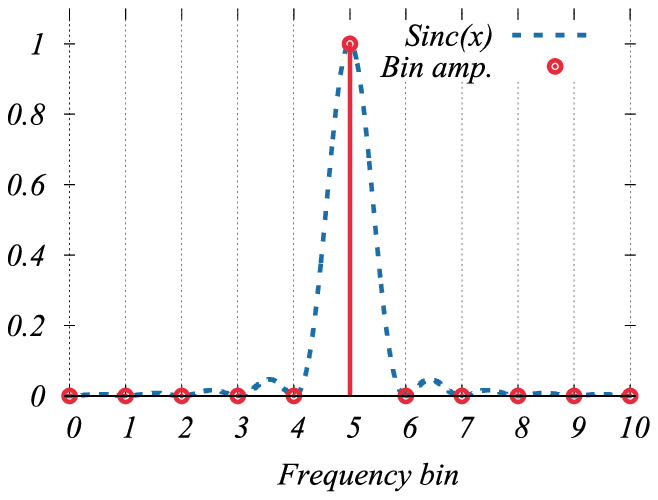}
	\end{minipage}%
	\begin{minipage}[t]{.45\textwidth}
		\centering
		\includegraphics[width=\linewidth]{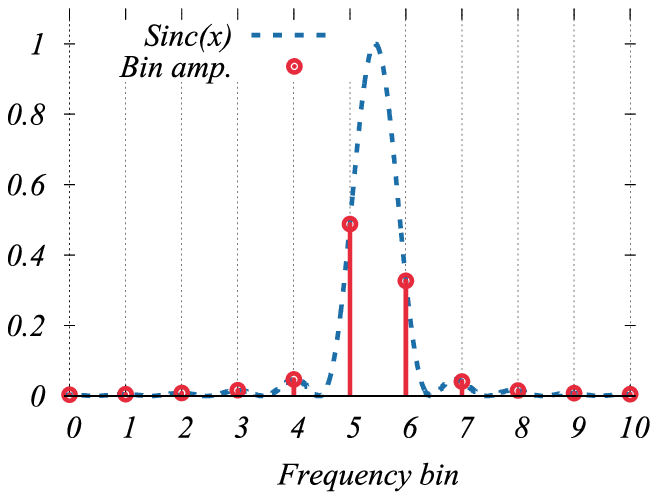}
	\end{minipage} \\ %
	\caption{Scalloping loss due to a discrete representation of frequencies. \label{fig:discrete_rep}}
\end{figure}

When the frequency we want to represent in the discrete spectrum does not coincide with the centre of any of the frequency bins, then the signal's power is distributed across all frequency bins -- the distribution follows a Sinc function. The consequences of the redistribution of power across all frequency bins is shown in Figure \ref{fig:discrete_rep}, where we see a noticeable decrease in amplitude for the frequency outside the centre of the frequency bin compared to the represented frequency at the centre of the frequency bin. The effect of decreased amplitude for off-centre frequencies is called a \textit{scalloping loss}. The scalloping loss can cause up to 36\% loss \citet{LYONS:UDSP} of the amplitude, assuming we know which bin to pick.

\begin{figure}[htp]
    \centering
	\begin{minipage}[t]{.45\textwidth}
		\centering
		\includegraphics[width=\linewidth]{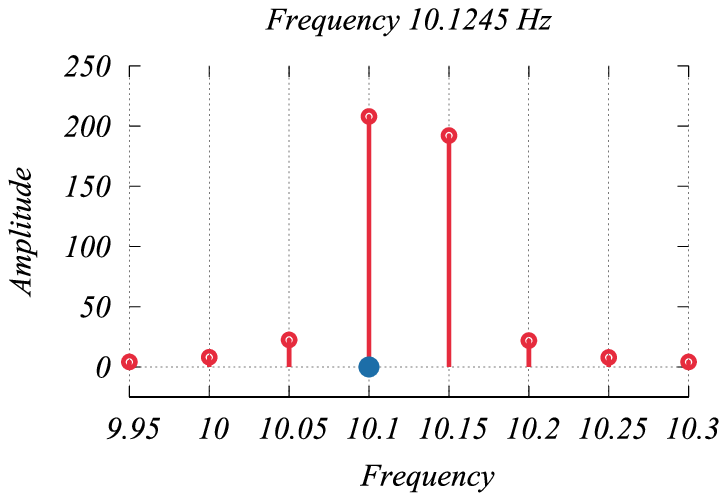}
	\end{minipage}%
	\begin{minipage}[t]{.45\textwidth}
		\centering
		\includegraphics[width=\linewidth]{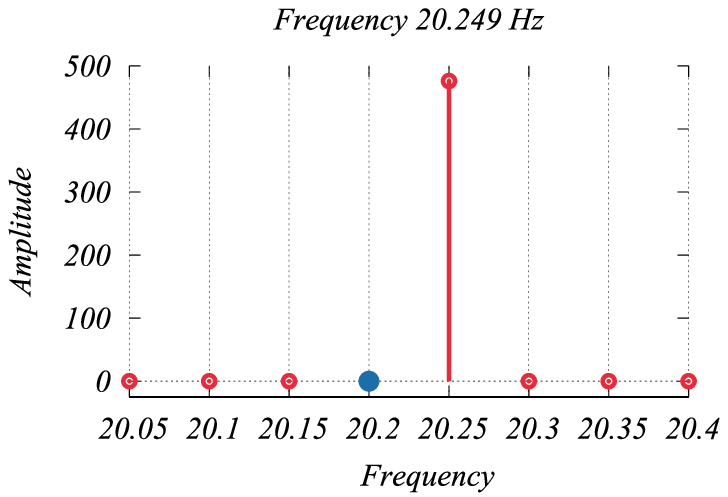}
	\end{minipage}%
	\\
	\begin{minipage}[t]{.45\textwidth}
		\centering
		\includegraphics[width=\linewidth]{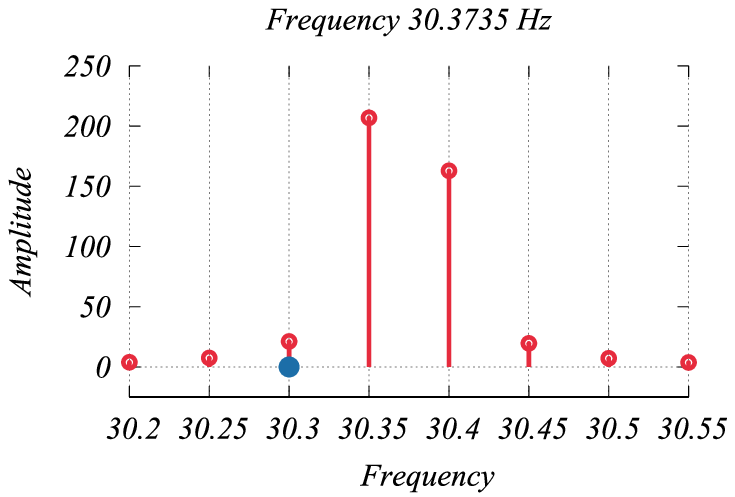}
	\end{minipage}%
	\begin{minipage}[t]{.45\textwidth}
		\centering
		\includegraphics[width=\linewidth]{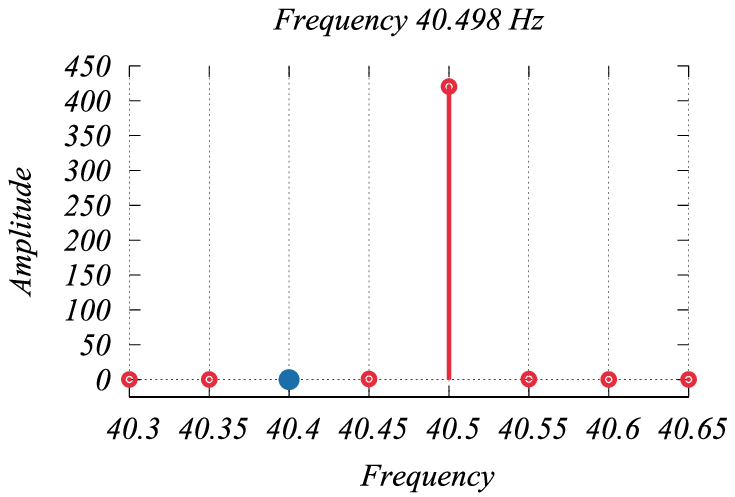}
	\end{minipage}\\
	\caption{Signal of the pulsar with fundamental frequency $f=10.1245 \mathrm{Hz}$ after Fourier transformation. Individual plots shown are fundamental frequency and the first three harmonics. The integer multiple of the fundamental frequency bin ($f_b=10.1 \mathrm{Hz}$) is emphasised by the filled blue point. \label{fig:shift}}
\end{figure}

\subsection{Peak drift}
The problem of picking the correct harmonic bin for summation is that the representation of the higher harmonic frequencies is, in the discrete spectrum, based on the integer multiple of the pulsar's true frequency $f_h = h\fp$ and not on the fundamental frequency $\ff$. Therefore, there might be a frequency bin that more closely represents the higher harmonic $f_h$. The frequency of this bin might be different from the integer multiple of the fundamental frequency bin $h\ff$, thus introducing \textit{drift} in the peak position at higher harmonics. The drift is illustrated in Figure \ref{fig:shift}, where the frequency $f=10.1245$ is chosen to be almost exactly in between two frequency bins. A filled blue point depicts the fundamental frequency bin and its integer multiples. We see that at higher harmonics, a mere multiple of the fundamental frequency bin would miss the peak and would result in no increase to the signal-to-noise ratio of the pulsar. 

\begin{figure}[htp]
	\centering
	\includegraphics[width=0.8\linewidth]{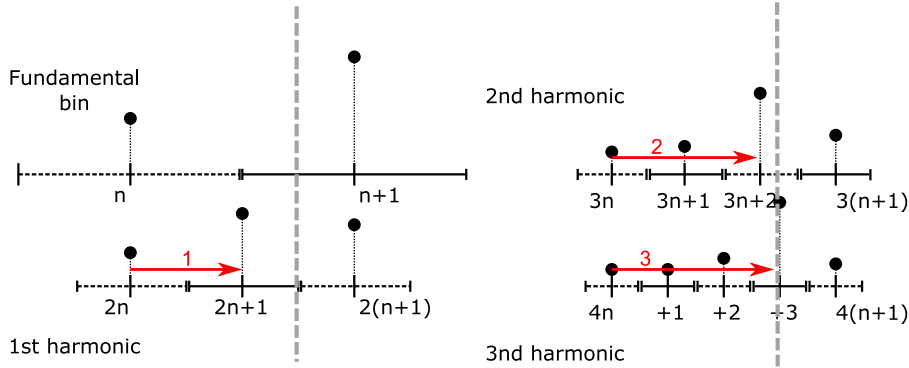}
	\caption{Frequency bin drift, when interpreted from one bin centre to another, is an increasing function. The drift and its value are in red. The grey dashed vertical line represents the true frequency value and its integer multiples at higher harmonics.
    \label{fig:bindrift}}
\end{figure}
	
In this work, we have adopted an interpretation that the peak can drift from one centre of the frequency bin to another, rather than being a peak's distance from the centre of one frequency bin. Therefore, we define peak drift as the distance from the fundamental frequency bin (or its integer multiple) to the peak at the same harmonic order. In our interpretation, the drift is an increasing function, while the value of deviation from the frequency bin $n+1$ centre could also decrease. Our chosen interpretation is depicted in Figure~\ref{fig:bindrift}, where the true frequency is outside the fundamental frequency bin $n$ and lies in the first half of the frequency bin $n+1$. As we sum higher harmonics, the peak (true pulsar frequency) position is more precisely represented. This is because, in our interpretation, higher harmonics will have a higher resolution in frequency because the interval between bins $n$ and $n+1$ is mapped into more and more bins, which lay between bins $hn$ and $h(n+1)$, as the harmonic order $h$ increases. 

\begin{figure}[htp]
	\centering
	\includegraphics[width=0.70\linewidth]{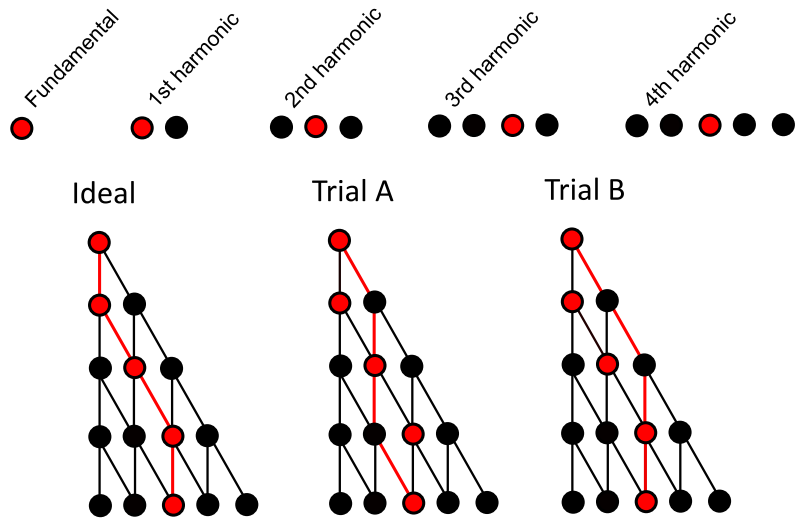}
	\caption{One potential partial sum of the harmonic sum algorithm for a fixed fundamental frequency bin could be interpreted as a path in a tree where the layers are made out of frequency bins at higher harmonics. Each path would then be defined only by the drift value at harmonic order $h$. \label{fig:drifttree}}
\end{figure}

Since the drift is either increasing by one or stays the same, we can picture it as a tree of nodes, as shown in Figure~\ref{fig:drifttree}, where there is an ideal path and two different trials. With this model, each harmonic sum for a fixed fundamental frequency is one path through the tree. 

The \textit{maximum drift} a peak undergoes depends on the harmonic order which is being summed, and it is given as $d_\mathrm{m}(h)=h-1$. This is the number of bins between two consecutive fundamental frequency bins at the same harmonic order. The actual drift a peak acquires depends on the pulsar's frequency and increases with the difference between the fundamental frequency $\ff$ and the true frequency of the pulsar $\fp$.


\section{Algorithm and Implementation}
\label{sec:algorithms}
We have developed a new harmonic sum algorithm based on the greedy approach. The algorithm prefers short term SNR gain rather than the optimal partial sum, which could give the higher SNR gain for a given fundamental bin. This algorithm is then compared with the established harmonic sum algorithm, which is used in the PRESTO software package \citet{Presto}. The second harmonic sum algorithm used for comparison is the simple harmonic sum which ignores the drift and sums only integer multiples of the fundamental bin. We have implemented GPU versions of all of these harmonic sum algorithms, and we compared these algorithms in performance and sensitivity.

\begin{figure}[htp]
    \centering
	\begin{minipage}[t]{.35\textwidth}
		\centering
		\includegraphics[width=\linewidth]{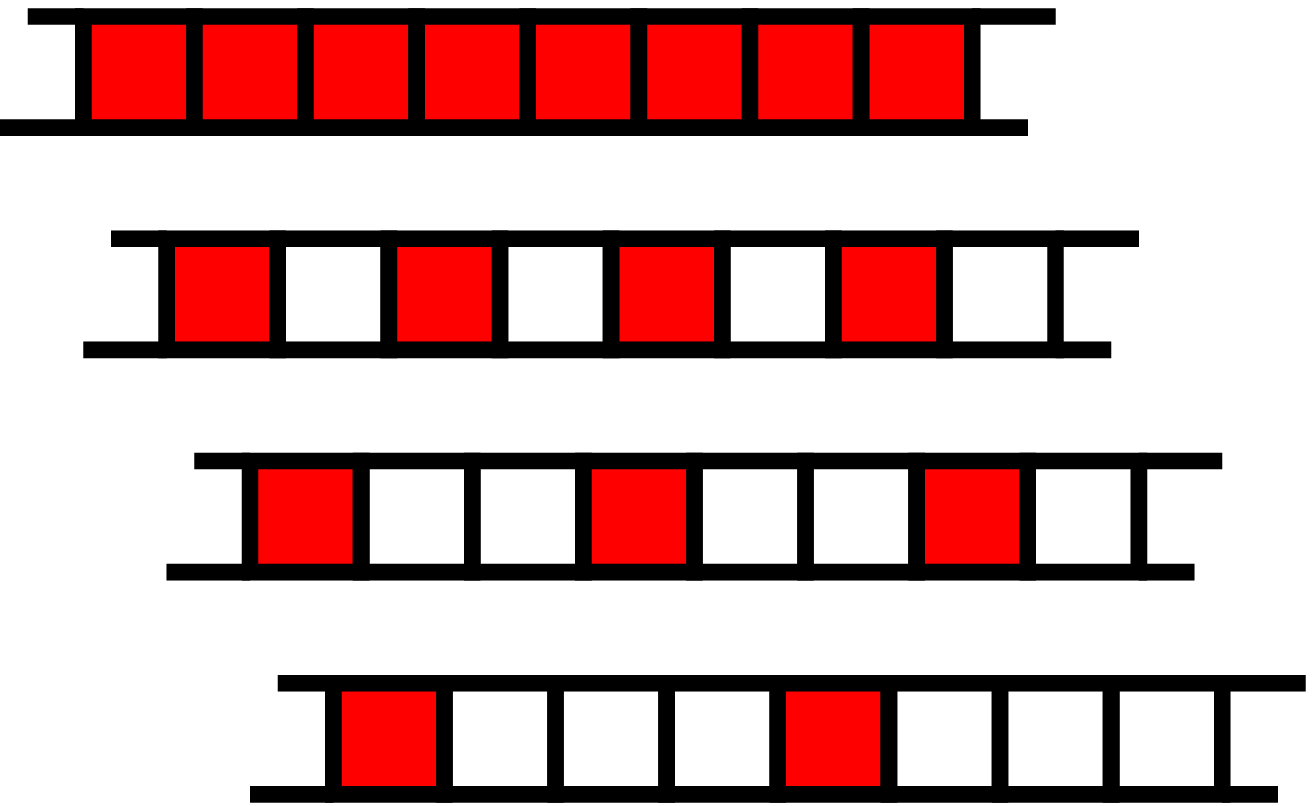}
	\end{minipage} %
	\begin{minipage}[t]{.35\textwidth}
		\centering
		\includegraphics[width=\linewidth]{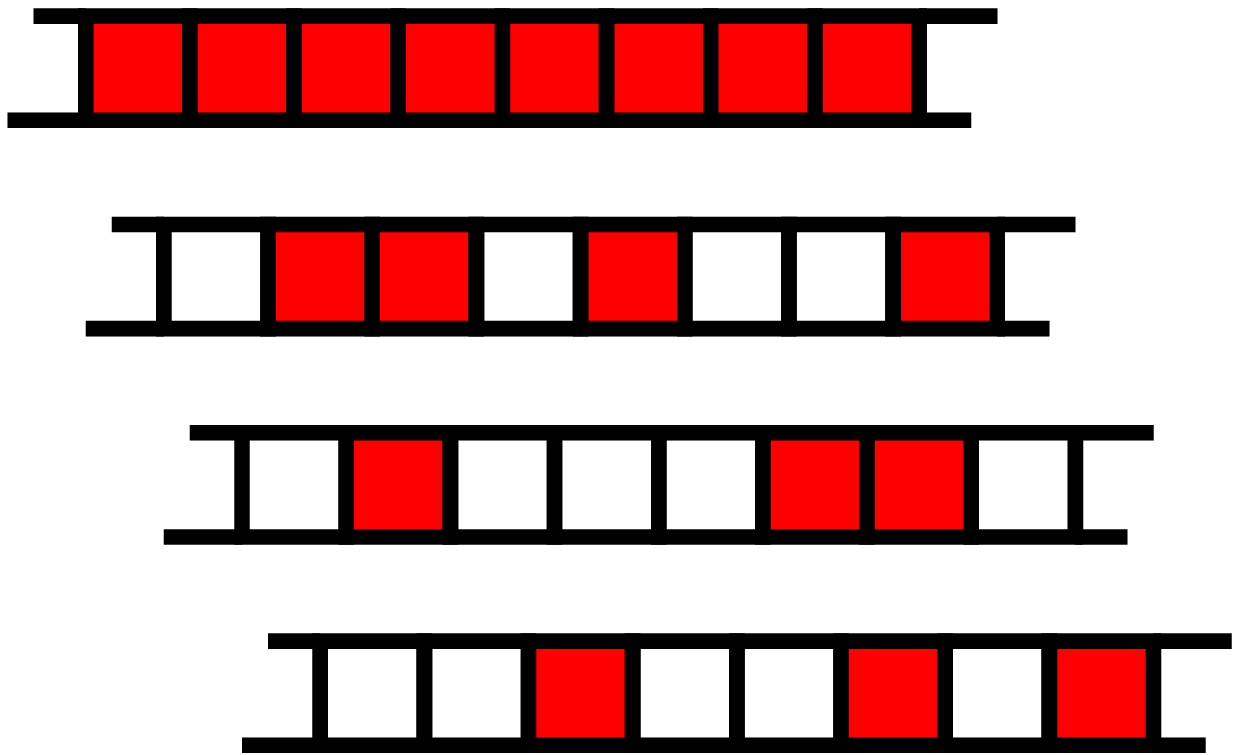}
	\end{minipage} \\%
	\caption{Stride in the memory access pattern for the harmonic sum without drift on the left and including drift on the right. \label{fig:simple_access}}
\end{figure}

Any harmonic sum algorithm must solve two main problems. The first is an unfavourable memory access pattern, and the second is the number of possible partial sums explored per a single fundamental bin. 

The poor memory access pattern is due to the distance between integer multiples of two consecutive fundamental bins. This distance increases with the harmonic order, which is being added to the partial sum starting with one up to the maximum harmonic order. In other words, we access the data with an increasing stride. Figure~\ref{fig:simple_access} (left) illustrates such a memory access pattern that, in addition, ignores the peak's drift. Taking peak drift into account, we add randomness into the memory access pattern, as shown in Figure~\ref{fig:simple_access} (right). The stride in any memory accesses represents a problem because it significantly reduces data utilisation per cacheline\footnote{A cacheline refers to a block of data that is spatially next to each other and transferred together in one memory transfer. This is due to memory architecture.} unless this is mitigated by CPU/GPU cache.

The second problem faced by the harmonic sum algorithm is how to limit the number of partial sums performed. As the peak drift is unknown, to get the highest SNR, we need to explore a number of partial sums containing different elements to estimate the value of the drift. However, the calculation of these various partial sums can be computationally expensive. Therefore the harmonic sum algorithm must prioritise which partial sums to use.  

In the following sections, we will describe our new Greedy harmonic sum algorithm as well as other harmonic sum algorithms that are used for comparison.


\subsection{Simple harmonic sum}
The simple way to perform the harmonic sum is to ignore the peak's drift and sum only the values located at the integer multiple of the fundamental frequency bin. This approach has the potential to be fast, but it sacrifices SNR gain for performance. The simple harmonic sum can be summarised by the equation
\begin{equation}
    X[n]=\sum_{h=1}^{H}x\left[hn\right]\,,
\end{equation}
where $n$ is the fundamental frequency bin and $h$ is the harmonic order. 
The simple harmonic sum serves as a bottom line for other harmonic sum algorithms regarding SNR gain. 

\subsubsection{Implementation}
In our GPU implementation of the simple harmonic sum, each GPU thread calculates the harmonic sum for one fundamental bin. The threads that are next to each other process fundamental bins in sequence. Since the data for higher harmonics are accessed with an increasing stride, as discussed above and shown in Figure~\ref{fig:simple_access}, this decreases cacheline utilisation. When reading fundamental bins, a cacheline is fully utilised. In contrast, for higher harmonics, it is utilised to $1/h$, where $h$ is the harmonic order used. Cacheline utilisation has the potential to decrease performance if not mitigated by caches. In the case of the simple harmonic sum, which does not suffer from random access to memory introduced by the drift, we can improve the access pattern if we transpose the data. Instead of having fundamental bins from a single time-series spatially close to each other, we change the layout of the data to put the same bin from different time-series next to each other. The advantage of this transformation is that when reading data for higher harmonic orders, we read data that are contiguous in memory. That way, threads process the same fundamental bin but from different time-series. It also means we have to pay the price of transposing data, however.

During our investigation, we discovered that the transpose of the data is not beneficial on the newer generation of NVIDIA GPUs since they have a write-through cache. The older generations of NVIDIA GPU benefit more from such pre-processing.


\subsection{PRESTO harmonic sum}
The harmonic sum algorithm used in the PRESTO software package is based on the following formula
\begin{equation}
X[n]=\sum_{h=1}^{H}x\left[\frac{h}{H}n\right]\,,
\end{equation}
where as before $H$ is the number of harmonics to sum, and $x$ is the input time-series. The index $n$ in the case of the PRESTO algorithm is not the fundamental bin but the last bin to be summed in a harmonic series.    

\begin{algorithm}
  \SetAlgoLined
    \SetKw{KwBy}{by}
    \SetKwFunction{snr}{CalculateSNR}
    \SetKwFunction{compare}{CompareSNR}
    \SetKwFunction{round}{Round}
    \SetKwFunction{msd}{GetMeanStdev}
    \SetKwFunction{array}{array}
    \SetKwData{sd}{SumDown}
    \SetKwData{sl}{SumLeft}
      \array $\mu[H_\mathrm{max}], \sigma[H_\mathrm{max}]$\;
      $\left\{\mu,\sigma\right\}$ = \msd()\;
      Sum = 0\;
      $\mathrm{SNR}_\mathrm{max}$ = 0\;
      $h_\mathrm{m}$ = 0\;
      \For{$H=1$ \KwTo $H_\mathrm{max}$ \KwBy $H=2H$ }{
        \emph{Fundamental frequency bin}\;
        P = \round$(1/Hn)$\;
        Sum = $x$[P]\;
        \For{$h=2$ \KwTo $H$ \KwBy $h=h+2$}{
          P = \round$(h/Hn)$\;
          Sum = Sum + $x$[P]\;
          $\mathrm{SNR}$ = \snr$(\mathrm{Sum},\mu[h],\sigma[h])$\;
          $\left\{\mathrm{SNR}_\mathrm{max}, h_\mathrm{m}\right\}$ = \compare$(\mathrm{SNR},h,\mathrm{SNR}_\mathrm{max},h_\mathrm{m})$\;
        }
        $\mathrm{SNR}_o = \mathrm{SNR}_\mathrm{max}$\;
        $h_\mathrm{o} = h_\mathrm{m}$
      }
   \caption{Pseudo-code for the PRESTO harmonic sum for a bin $n$. A series of harmonic sums of increasing harmonic order $H$ is performed on the input data $x$ up to a maximum harmonic order $H_\mathrm{max}$. Output is the maximum SNR found $\mathrm{SNR}_\mathrm{max}$ and related harmonic order $h_\mathrm{m}$. To calculate SNR, mean $\mu[h]$ and standard deviation $\sigma[h]$ is used.}
  \label{alg:PRESTO}
\end{algorithm}

\begin{figure}[htp]
	\centering
	\includegraphics[width=0.75\linewidth]{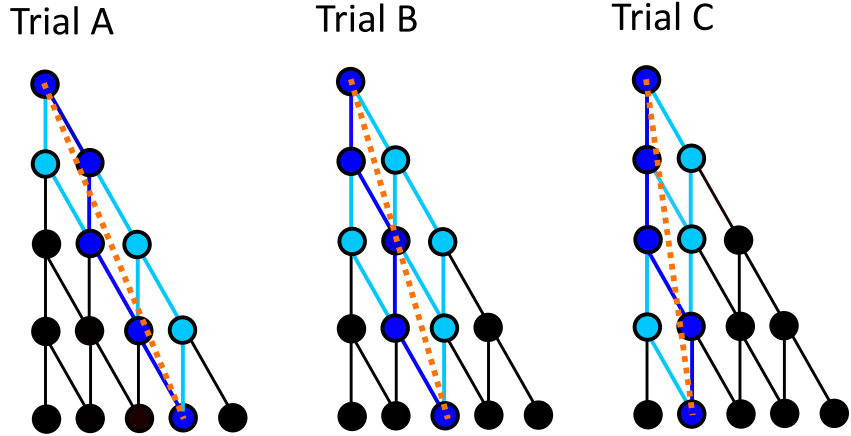}
	\caption{PRESTO chooses a path based on the last element of the sum.  \label{fig:PRESTOdrift}}
\end{figure}

By starting from the last element, some paths are eliminated, the remaining paths are further reduced by selecting only one path by rounding bin location $(h/H)n$ to the nearest integer. Paths are chosen for the same fundamental, but different end elements of the harmonic sum are shown in Figure~\ref{fig:PRESTOdrift}, where the chosen path is shown in dark blue, and all the possible paths leading to that element are shown in light blue. The pseudo-code for the PRESTO harmonic sum algorithm can be found in Algorithm~\ref{alg:PRESTO}. 

The advantage of this algorithm is that it covers most of the drift possibilities and it is deterministic. The disadvantage of this approach is that the last element from which the algorithm begins, acts as the last element for multiple fundamental bins but at different harmonic orders. For example, bin 800 is the last element for the fundamental frequency bin 100 at $H=8$ or for fundamental frequency bin 200 at $H=4$. Furthermore, the set of elements that need to be summed differs by the fundamental bin. Therefore, we have to perform an additional loop (the first loop in Algorithm~\ref{alg:PRESTO}) to cover these fundamental bins and thus increase the complexity of the algorithm to $\mathcal{O}(H^2)$. These harmonic sums may include different elements depending on rounding, making any data reuse difficult. 

The PRESTO algorithm for each bin $n$ performs all partial sums that are power-of-two long $H_P=2^i$, that is, for $H=32$ partial sums performed would be $H_P=\left\{2, 4, 8, 16, 32\right\}$. Further, PRESTO sums only even harmonics. These choices lower the complexity of the algorithm to $\mathcal{O}(H\log_2(H))$, but at the cost of sensitivity.

There are two ways to decrease sensitivity loss and thus improve the SNR of the PRESTO harmonic sum. Improvements however, come at the expense of algorithm performance. The first improvement is to sum elements of all higher harmonics, which we will call $\mathrm{PRESTO}_\mathrm{all}$. The second is to perform the harmonic sum for all harmonic orders in addition to summing all higher harmonics. We will call this algorithm PRESTO+.

We have used the GPU implementation of the PRESTO harmonic sum and PRESTO+ harmonic sum as a standard to which we compare the greedy harmonic sum algorithm. 



\subsection{Greedy harmonic sum}
    The greedy algorithm tracks the peak position and updates the drift value if the peak shifts. The greedy algorithm is listed in Algorithm~\ref{alg:greedy}. At each new harmonic order, the algorithm loads values of two consecutive frequency bins. The first frequency bin index $m$ is 
    \begin{equation}
    m = hn+d\,,
    \end{equation}
    where $n$ is the fundamental bin index, $h$ is the harmonic order, and $d$ is the value of the drift. Values of both bins are compared, and if the value of the second frequency bin is higher, the peak has shifted and drift is increased by one.
    
    \begin{algorithm}
  \SetAlgoLined
    \SetKwBlock{getbins}{GetBinValue(bin index: $p$) begin}{end}
    \SetKwBlock{hrms}{HarmonicSum begin}{end}
    \SetKwFunction{snr}{CalculateSNR}
    \SetKwFunction{interpolate}{Interpolate}
    \SetKwFunction{compare}{CompareSNR}
    \SetKwFunction{msd}{GetMeanStdev}
    \SetKwFunction{binvalue}{GetBinValue}
    \SetKwFunction{array}{array}
    \SetKwData{scm}{enable\_scalloping\_loss\_mitigation}
    \SetKwData{sl}{SumLeft}
        \array $\mu[H], \sigma[H]$\;
        $\left\{\mu,\sigma\right\}$ = \msd()\;

        \hrms{
            \For{$n=1$ \KwTo $N$}{
                sum = 0\;
                d = 0\;
                $\mathrm{SNR}_\mathrm{max}$ = 0\;
                $h_\mathrm{m}$ = 0\;
                \emph{fundamental bin}\;
                $x_\mathrm{d}$ = $x[n]$\;
                $x_\mathrm{(d+1)}$ = $x[n+1]$\;
                \uIf{$x_\mathrm{(d+1)}$ $>$ $x_\mathrm{d}$}{
                    d++\;
                    sum = sum + $x_\mathrm{(d+1)}$\;
                }
                \Else{
                    sum = sum + $x_\mathrm{d}$\;
                }
                $\mathrm{SNR}_\mathrm{max}$ = \snr(sum,$\mu[0]$,$\sigma[0]$)\;
                \emph{higher harmonics}\;
                \For{$h=1$ \KwTo $H$}{
                    $x_\mathrm{d}$ = $x[hn+d]$\;
                    $x_\mathrm{(d+1)}$ = $x[hn+d + 1]$\;
                    \uIf{$x_\mathrm{(d+1)}$ $>$ $x_\mathrm{d}$}{
                        d++\;
                        sum = sum + $x_\mathrm{(d+1)}$\;
                    }
                    \Else{
                        sum = sum + $x_\mathrm{d}$\;
                    }
                    $\mathrm{SNR}$ = \snr(sum,$\mu[h]$,$\sigma[h]$)\;
                    $\left\{\mathrm{SNR}_\mathrm{max}, h_\mathrm{m}\right\}$ = \compare($\mathrm{SNR}$, $h$, $\mathrm{SNR}_\mathrm{max}$, $h_\mathrm{m}$)\;
                }
                $\mathrm{SNR}_o[n] = \mathrm{SNR}_\mathrm{max}$\;
                $h_\mathrm{o} = h_\mathrm{m}$
            }
        }
   \caption{Pseudo-code for the greedy harmonic sum. Where $H$ is maximum harmonic order summed, $x$ is input data in the frequency domain, $\mathrm{SNR}_\mathrm{max}$ is maximum SNR found, $h_\mathrm{m}$ is related harmonic order, and $\mu[h]$, $\sigma[h]$ is mean and standard deviation respectively for partial sum of $h$ elements. The algorithm's output is the maximum SNR found $\mathrm{SNR}_o[n]$ for each relevant fundamental frequency bin and related harmonic order $h_\mathrm{o}$. The variable drift keeps track of the peak drift in the number of frequency bins.}
  \label{alg:greedy}
\end{algorithm}

    The algorithm for following the peak position as harmonic order increases is inspired by the binary search algorithm. The binary search algorithm finds a number in a sorted array of numbers by splitting the array into two sub-arrays. Then (for an ascending array), the number in the middle is compared to the number we are searching for, and if the number we are searching for is smaller, we continue the search in the first half of the array. If not, the second half of the array is used for further searches. The total number of sub-arrays is doubled with each iteration, as each sub-array is split in two. This allows us to contain the number we are searching for in ever-smaller sub-arrays until it is found.

    \begin{figure}[htp]
    	\centering
    	\includegraphics[width=0.75\linewidth]{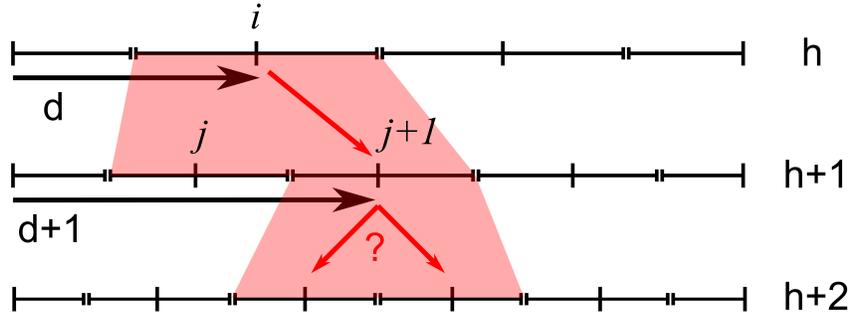}
    	\caption{Localisation of the peak using a binary search like algorithm. \label{fig:binmapping}}
    \end{figure}

    In contrast to the binary search method, the number of sub-intervals that divide the interval between two consecutive fundamental frequency bins (fundamental interval) is not doubled at each harmonic order but only increased by one to $n+1$ dividing sub-intervals. The situation is depicted in Figure~\ref{fig:binmapping}. 
    
    Let's suppose that we want to use a similar algorithm to the binary search for peak position at higher harmonic orders. In this case, we must show that a frequency bin $i$ with drift $d$ and width $w$ at harmonic order $h$ is contained in two consecutive frequency bins $j$ and $j+1$ at harmonic order $h+1$. 
    That is, the frequency $f_{i,b} = f_i - w/2$ at the beginning of the bin $i$, mapped to a higher harmonic order $h+1$, must be higher than the frequency $f_{j,b} = f_j -w/2$ at the beginning of the frequency bin $j$:
    \begin{equation}
    f_{j,b} \geq \frac{h+1}{h}f_{i,b}\,,
    \label{eqa:greedycond1}
    \end{equation}
    where $(h+1)/h$ maps frequencies from harmonic order $h$ to the frequency range of the harmonic order $h+1$. Substituting $f_i = hf_f+dw-w/2$ and $f_j = (h+1)f_f+dw-w/2$ into equation~(\ref{eqa:greedycond1}) gives condition
    \begin{equation}
    2d-1 \geq 0\,,
    \label{eqa:greedycond1_b}
    \end{equation}
    which holds for $d>0$. The bin at $d=0$ is at the beginning of the fundamental interval where we have only the second half of the bin therefore this rule cannot be applied. However all bins with $d=0$ begin at integer multiple of the fundamental bin therefore the condition is met.

    The second condition is that the end of the bin $i$ must be less than, or equal to, the end of the second bin $j+1$. That is
    \begin{equation}
       \frac{h+1}{h}f_{i,e} \leq f_{(j+1),e}\,,
        \label{eqa:greedycond2}
    \end{equation}
    where $f_{i,e}=hf_f + dw + w/2$ is the frequency at the end of the frequency bin $i$ and $f_{(j+1),e}=(h+1)f_f + (d+1)w + w/2$ is the frequency at the end of the bin $j+1$. After substitution we arrive at the condition
    \begin{equation}
        2d+1 \leq 2h\,.
        \label{eqa:greedycond2_b}
    \end{equation}
    As the maximum value of the drift is $d=h-1$, the inequality~(\ref{eqa:greedycond2_b}) holds for any value $d>0$.
    
    Hence, we can track the position of the peak as the drift $d$ changes with increasing harmonic order $h$. 
    
\begin{figure}[htp]
    	\centering
    	\includegraphics[width=\linewidth]{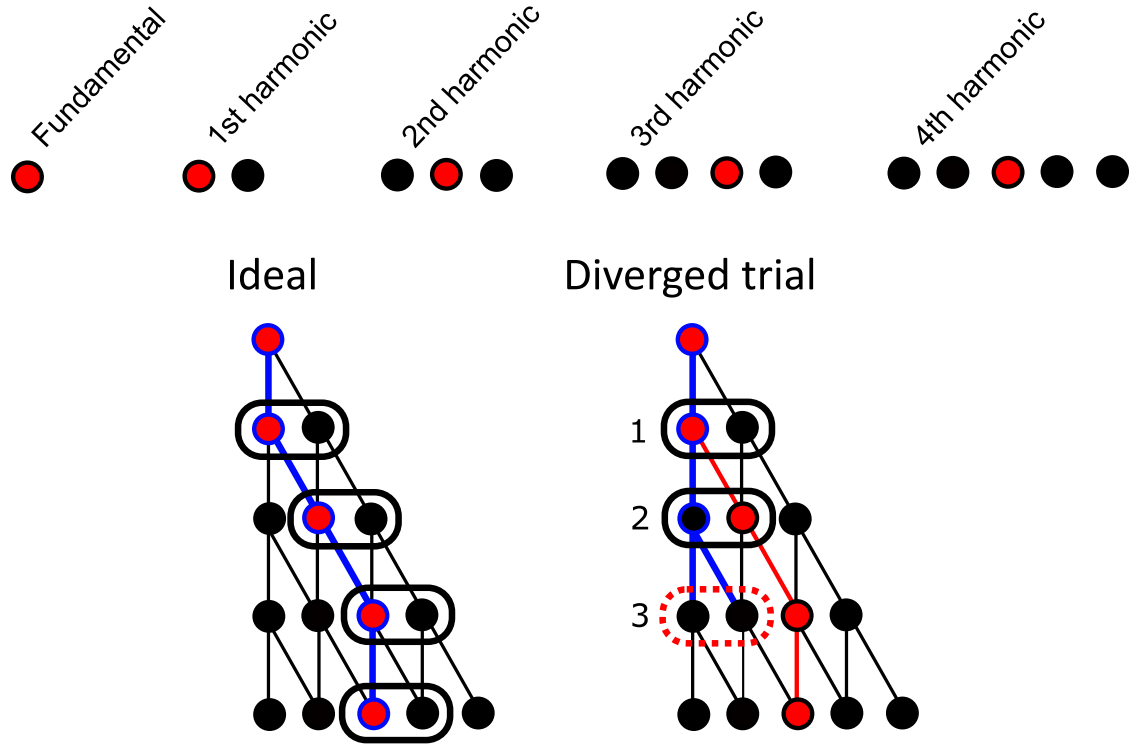}
    	\caption{Greedy algorithm. Decision points are shown as rounded boxes. \label{fig:greedyalg}}
    \end{figure}

    The greedy algorithm evaluates which element to sum based on short term SNR gain. Each rounded box in Figure~\ref{fig:greedyalg} shows a decision that chooses the next direction of the summation. However, a single element containing a high noise value can potentially divert the greedy algorithm to a path which, in the long term, sums to a lower SNR than it otherwise would have if the algorithm followed the correct path. The divergent path is shown in Figure~\ref{fig:greedyalg} on the right.
    
    To increase the recovered SNR further, we can apply the same principle of selecting the bin with the highest value for the fundamental frequency bin. Since drift is defined from one bin centre to the next, it is possible that the true frequency is closer to the frequency bin $n+1$, which is next to the fundamental frequency bin $n$. Bin $n+1$ would then have a higher amplitude than the fundamental frequency bin $n$. However, as the true frequency is lower in the centre of the bin $n+1$, the higher harmonics are contained between integer multiples of bins $n$ and $n+1$. This increases recovered SNR.


\section{Results}
\label{sec:results}
    The comparison between our Greedy Harmonic sum algorithm and both PRESTO harmonic sum algorithms are presented in this section. We will compare the implemented harmonic sum algorithms ability to detect pulsar signals and their compute performance. To compare the ability to detect pulsars, we use the recovered SNR (RSNR) that is the SNR returned by the harmonic sum algorithm. We distinguish recovered SNR from the injected SNR of the signal as the RSNR depends on other steps, chiefly on calculating the mean and standard deviation. Therefore the RSNR may deviate from the injected SNR of the signal. However, all algorithms use the same pre-processing steps (mean and standard deviation calculation, power spectrum calculation, and interbinning). Thus the presented RSNR values are comparable between algorithms. The SNR is calculated using equation~\ref{eqa:SNR}. Only the highest SNR value is returned for each element $n$.

    Besides comparing absolute values of RSNR, we will also compare the sensitivity loss incurred when the pulsar frequency is between the centre of frequency bins and the dependency of the RSNR on pulsar frequency. For these comparisons, we will use artificial data produced using the von Mises distribution. Following this, we will compare all algorithms using synthetic data with white noise included. Lastly, we will test both algorithms integrated into the AstroAccelerate software package, where we will compare the result from real pulsar data from the GHRSS survey \citet{Bhattacharyya2016:GHRSS} using the Giant Metrewave Radio Telescope (GMRT).

    The performance of the harmonic sum algorithm will be measured in the number of processed fundamental frequency bins per second. Processing a fundamental frequency bin means constructing a partial sum, or sums, and returning the SNR value.

\subsection{Artificial data without noise}
    To create an artificial pulsar, we have followed \citet{Ransom2002:FDAS} which used a modified von Mises distribution. The modified von Mises distribution is given by
    \begin{equation}
        f(\kappa,t) = a\frac{ e^{\kappa\cos\left( 2 \pi f_r t + \varphi \right)} - e^{-\kappa} }{I_0(\kappa) - e^{-\kappa}}\,,
        \label{eqa:vonMises}
    \end{equation}
    where $a$ is amplitude, $f_r$ is the pulsar's frequency, $0\leq t \leq T$ is time, $T$ is the time of the observation, $\varphi$ is the phase, and $I_0$ is the modified Bessel function of zeroth order. The sampling time used was $t_s=64\mu\mathrm{s}$ The shape parameter $\kappa$ can be used to change the full width at half-maximum (FWHM), therefore the duty cycle of the pulsar which affects the number of significant harmonics present in the power spectra. More details can be found in \citet{Ransom2002:FDAS}.  This way, we can create artificial pulsar data that can be used to test and evaluate the harmonic sum algorithm. 

\subsubsection{Sensitivity loss}
    The sensitivity loss comes from two sources that are connected. The first is the drift of the peak. The second is the scalloping loss. In order to have a non-zero drift, the true frequency of the pulsar must be outside of the centre of the frequency bin, which also means that the signal will suffer from scalloping loss. 
    
    To evaluate the sensitivity loss of a harmonic sum algorithm, we have created a set of time-series with increasing true pulsar frequency $\fp$ from one centre of the frequency bin $n$ to the next centre of the frequency bin $n+1$. We have set the observation length to $T=20s$ with sampling time $t_s=64\mu s$, thus the width of the frequency bin is $W_f=0.05\,\mathrm{Hz}$. The frequency bin width was then divided into 50 steps. To determine the sensitivity loss of an algorithm, we have normalised the RSNR value returned by an algorithm, by the value of RSNR returned by the same algorithm at the centre of the frequency bin. That is, the sensitivity loss at the pulsar frequency $f$ is
    \begin{equation}
        l(f) = \frac{\mathrm{RSNR}(f)}{\mathrm{RSNR}(\ff)}\,.
        \label{eqa:sensitivityloss}
    \end{equation}
    Two algorithms may have similar sensitivity loss but different RSNR and thus different probabilities of detecting a pulsar.
    
\begin{figure}[htp]
        \centering
    	\begin{minipage}[t]{.48\textwidth}
    		\centering
    		\includegraphics[width=\linewidth]{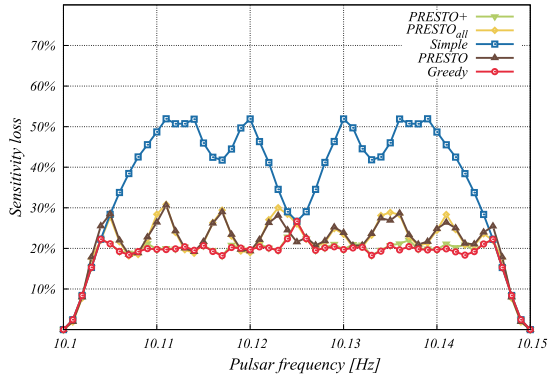}
    	\end{minipage} %
    	\caption{Sensitivity loss of different harmonic sum algorithms. \label{fig:sensitivityloss}}
    \end{figure}

    The value of the drift increases with pulsar frequency, and scalloping loss increases until the border of two bins is reached, where it starts to diminish upto the point of having no effect at the centre of the next frequency bin. If the tracking of the drift in a harmonic sum is correct, we should see a symmetrical sensitivity loss around the border of the two bins. The comparison of the two versions of the PRESTO algorithms, simple harmonic sum and Greedy harmonic sum algorithm in sensitivity loss is shown in Figure~\ref{fig:sensitivityloss}.

\subsubsection{Recovered SNR}
\begin{figure}[htp]
        \centering
    	\begin{minipage}[t]{.48\textwidth}
    		\centering
    		\includegraphics[width=\linewidth]{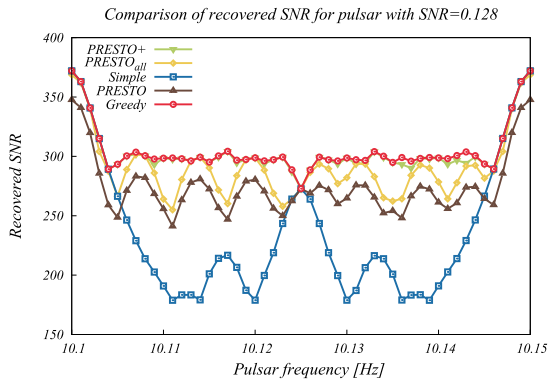}
    	\end{minipage} 
    	\caption{Recovered SNR by different harmonic sum algorithms. \label{fig:recoveredSNR}}
    \end{figure}

    The RSNR returned by an harmonic sum algorithm depends also on the number of harmonic orders summed. The one of the two differences between PRESTO, PRESTO$_\mathrm{all}$, PRESTO+ and the Greedy harmonic sum algorithm is the number of harmonic orders summed. The recovered SNR for different algorithms is shown in Figure~\ref{fig:recoveredSNR}. This is for $H=32$.

\begin{figure}[htp]
       \centering
    	\begin{minipage}[t]{.49\textwidth}
    		\centering
    		\includegraphics[width=\linewidth]{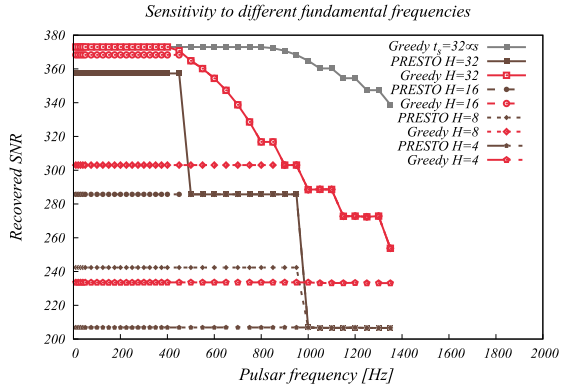}
    	\end{minipage}%
    	\caption{Sensitivity loss of different fundamental frequencies and different maximum harmonic orders $H$. \label{fig:difffundamentalsRSNR}}
    \end{figure}

    To show the dependence of RSNR on the different fundamental frequencies and the maximum number of harmonics summed, we can use an average RSNR. To calculate the average RSNR, we calculate a mean RSNR from all RSNR returned by a harmonic sum algorithm as the pulsar frequency changes from one centre of the frequency bin to the next. That is, a mean RSNR of all RSNRs shown in Figure~\ref{fig:recoveredSNR} for a given algorithm represents one average RSNR value. The average RSNR for different fundamental frequency bins and maximum harmonics summed is shown in Figure~\ref{fig:difffundamentalsRSNR}.

    The average RSNR decreases from the fundamental frequency $f=400\mathrm{Hz}$ because of a poor representation of the pulsar signal by the data. For sampling time $t_\mathrm{s}=64\mu s$, the pulse width is below two samples wide. Thus, in addition to the other effects discussed above, we have introduced more signal loss by undersampling the pulsar signal. For comparison, we have added a time-series with sampling time $t_\mathrm{s}=32\mu s$ that extends beyond $f=400\mathrm{Hz}$ and starts to decline at $f=800\mathrm{Hz}$ as expected. The artificial pulsar had about 30 significant harmonics and duty cycle of 4\%.

\subsection{Processing of observed data}
We have implemented Greedy, PRESTO and PRESTO+ harmonic sum algorithms into a GPU accelerated software package called AstroAccelerate \citet{AstroAccelerate}. AstroAccelerate is used for real-time data processing of time-domain radio astronomy data. We have tested and compared the performance of the greedy harmonic sum against PRESTO algorithms (PRESTO and  PRESTO+) using real GHRSS survey observations with the GMRT. The GHRSS (GMRT High Resolution Southern Sky) survey is an off-Galactic plane survey for pulsars and fast transients at 300$-$500 MHz using the GMRT already having 26 discoveries from CPU (using PRESTO, RIPTIDE \citet{2020MNRAS.497.4654M}) and GPU (using AstroAccelerate) processing. The GHRSS survey produces filterbank data with 4096 spectral channels sampled at 81.92 $\mu$s time-resolution. We have tested the harmonic sum algorithms using the in-beam pulsars detected in the GHRSS survey distributed over a range of periods and SNR.  
The results are shown in Table~\ref{tab:GMRTrecoveredSNR}.

    \begin{table*}
	\caption{Recovered SNR by the AstroAccelerate for in-beam pulsars in GHRSS survey data.}
	\centering
	\begin{tabular}{lrrrr}
		\toprule
		\multirow{2}{*}[-2pt]{Pulsar} &\multirow{2}{*}[-2pt]{Period(ms)}  &\multicolumn{3}{c}{Harmonic sum} \\
		\cmidrule(r){3-5} & & Greedy & PRESTO & PRESTO+ \\
		\midrule
		
		J0835-4510          &89.39 & 12k & 11k ($- 10\%$)   & 12k ($\phantom{+ 00\%}$) \\
		
        J1328-4357 &532.66 & 25  & 24 ($- \:\:7\%$) & 29 ($+ 14\%$) \\
        
        J1337-4441 &1257.41 & 36  & 26    ($- 28\%$) & 36 ($\phantom{+ 00\%}$) \\
        
        J1517-4356 &650.82 & 134 & 97    ($- 28\%$) & 163   ($+ 21\%$) \\
        
        J1517-4356 &650.78 & 53  & 42    ($- 27\%$) & 55  ($+ \:\:4\%$) \\
        
        J1542-5034 &599.27 & 37  & 34 ($- \:\:9\%$) & 36 ($- \:\:4\%$) \\
        
        J1613-4714 &382.34 & 209 & 163 ($- 22\%$)   & 209 ($\phantom{+ 00\%}$) \\
        
        J1708-3426 &692.16 & 25  & 21 ($- 15\%$)    & 24  ($- \:\:8\%$) \\
        
        J1823-3106 &284.08 & 105 & 79 ($- 25\%$)    & 104 ($\phantom{+ 00\%}$) \\
        
        J1817-3618 &387.04 & 105 & 79 ($- 25\%$)    & 104 ($\phantom{+ 00\%}$) \\  
        
        J0418-4154 &757.15 & 14  & 12 ($- \:\:8\%$) & 15  ($+ 14\%$) \\
        
		\bottomrule
		\end{tabular}
	\label{tab:GMRTrecoveredSNR}
\end{table*}

\subsection{Compute Performance}

\begin{table}
	\caption{Performance of the implemented harmonic sum algorithms measured in processed fundamental bins per second [Gbin/s].}
	\centering
	\begin{tabular}{lrrr}
		\toprule
		\multirow{2}{*}[-2pt]{H} & \multicolumn{3}{c}{Harmonic sum [Gbin/s]} \\
		\cmidrule(r){2-4}  & Greedy & PRESTO & PRESTO+ \\
		\midrule
		32   &  8.64 &  9.64 ($+ 11\%$)    &  1.16 ($- 87\%$) \\
		
        16   & 14.70 & 15.24 ($+ \:\:4\%$) &  3.79 ($- 74\%$) \\
        
        8    & 26.10 & 23.38 ($- 10\%$)    & 11.03 ($- 58\%$) \\
        
        4    & 39.9  & 37.49 ($- \:\:6\%$) & 26.80 ($- 33\%$) \\
		\bottomrule
		\end{tabular}
	\label{tab:computeperformance}
\end{table}

Compute performance was measured using an NVIDIA V100 GPU with 32GB of GRAM and evaluated as the number of fundamental frequency bins processed per second, which is calculated as
\begin{equation}
    \mathrm{[Gbin/s]}=\frac{\mathrm{Total\: number\: of\: frequency\: bins}}{\mathrm{Execution\: time}}\,,
\end{equation}
where the execution time was measured using the NVIDIA profiler. The tests were performed using 1024 power spectra, each 131 thousand elements long. All implemented harmonic sum algorithms are capable of searching through longer power spectra. The compute performance of all implemented harmonic sum algorithms is listed in Table~\ref{tab:computeperformance}.


\section{Discussion}
\label{sec:discussion}
The PRESTO harmonic sum algorithm is deterministic, and it is oblivious to the data it processes. That is, it always sums the same elements (with respect to the fundamental bin) given the same input parameters. On the other hand, the greedy harmonic sum algorithm changes what elements to sum based on the data it is processing. The non-deterministic feature of the greedy harmonic sum may pose a problem.  

Before discussing the performance and ability to recover a signal of any harmonic sum algorithm, we should discuss how the result is obtained, as this plays a vital role in signal extraction. Even a deterministic algorithm has to choose a figure of merit on which to judge whenever a particular partial sum of elements is better than any other. In both algorithms (PRESTO and greedy), this figure of merit is the value of SNR, where higher is better. This means that without any other knowledge, even the deterministic algorithm would select a partial sum that gives the highest SNR, even if it is the wrong one, over a partial sum that would contain pulsar's signal. Thus any harmonic sum algorithm could be misled by the noise. The advantage of the deterministic algorithm lies in that it does not abandon the partial sum prematurely. 

A greedy approach can be misled by spurious noise into a different path from which it may not recover. As we will see, this may have happened in a few cases as some detections by PRESTO+ are higher than detections by the greedy harmonic sum despite the same signal loss and recovered SNR values in test with artificial data without noise.

The sensitivity loss is shown in Figure~\ref{fig:sensitivityloss}. The highest signal loss occurs for the simple harmonic sum where the signal loss can be as high as 50\%, which shows that tracking the drift is essential for recovering high RSNR.

The figure also shows three different PRESTO algorithms that differ in sensitivity loss and recovered SNR. Those algorithms are the PRESTO algorithm used in the PRESTO software package, which has the second highest signal loss.
The $\mathrm{PRESTO}_\mathrm{all}$ algorithm that is based on the PRESTO algorithm but sums all harmonics for given maximum harmonic order $H$ instead of only even harmonics.
And lastly, the PRESTO+ algorithm that adds all harmonics for given maximum harmonic order $H$ and in addition performs the harmonic sum where the maximum harmonic order is increasing by one (from 1 to $H_\mathrm{max}$) instead of performing the harmonic sum only for $H$ that is a power-of-two. 

We see that $\mathrm{PRESTO}_\mathrm{all}$ has almost the same sensitivity loss as the PRESTO algorithm, but it has lower compute performance; hence we have not included it in further comparisons. The PRESTO+ algorithm has the same sensitivity loss as a greedy algorithm.

The recovered SNR is shown in Figure~\ref{fig:recoveredSNR}, where we see that it mostly follows the sensitivity loss behaviours. The PRESTO algorithm has lower recovered SNR than the rest of the algorithms because it does not sum all harmonics and does not explore all partial sums in the way PRESTO+ does. We also see that for data without noise, the greedy harmonic sum and PRESTO+ harmonic sum algorithms have the same RSNR.

The results for in-beam pulsars from the GHRSS survey in presence of real telescope noise, RFIs shows that the Greedy harmonic sum always detected higher RSNR than the PRESTO harmonic sum. However, there are cases where PRESTO+ has higher RSNR. In these cases, the greedy algorithm may have followed the wrong path due to noise. However, the greedy algorithm provides up to 28\% higher RSNR values than the PRESTO harmonic sum while offering a similar compute performance.

The performance of all implemented harmonic sum algorithms is shown in Table~\ref{tab:computeperformance}, where higher is better. We see that the greedy harmonic sum has a similar compute performance to the PRESTO harmonic sum algorithm. The greedy harmonic sum is slower by 11\% at maximum harmonic order search $H=32$ and 6\% faster at $H=4$. The PRESTO+ algorithm is the slowest. Scaling in the number of power spectra is linear for all implemented harmonic sum algorithms. Algorithms also scale linearly in length of the power spectra. However, each algorithm scales differently with maximum harmonic orders $H$ searched. The greedy harmonic sum scales linearly $\mathcal{O}(H)$ as an increase in the maximum harmonic order by one is means increasing the number of steps taken by the algorithm by one. The PRESTO+ algorithm scales as $\mathcal{O}(H^2)$, which is a disadvantage especially prominent in the transition from $H=16$ to $H=32$. The PRESTO algorithm scales as $\mathcal{O}(H\log H)$ because the outer loop only goes through $H$, which are powers of two.

\section{Conclusions}
\label{sec:conclusions}
We have demonstrated that our new greedy harmonic sum algorithm has lower signal loss and better recovered signal-to-noise ratio (SNR) than the standard algorithm used in the PRESTO software package while achieving similar performance in the number of processed fundamental frequency bins per second. 

The signal loss for the pulsar frequencies that fall outside of the centre of the frequency bins is lower by as much as 10 percentage points. The greedy harmonic sum algorithm recovers on average 20\% higher SNR than the PRESTO harmonic sum algorithm. The range of recovered SNR was from 7\% up to 28\% higher for the greedy harmonic sum algorithm compared to the PRESTO harmonic sum algorithm.

In terms of the performance, the greedy harmonic sum algorithm is 6\% faster than the PRESTO harmonic sum algorithm for the maximum harmonic order search $H=4$ and $H=8$ and 11\% slower for harmonic order $H=32$.

We have also compared our greedy harmonic sum to an improved PRESTO harmonic sum algorithm that achieves the same signal loss values and recovered SNR. However, the greedy harmonic sum algorithm is $7\times$ faster for $H=32$ and $1.4\times$ faster for $H=4$.

\section*{Acknowledgements}
The authors acknowledge the support of the OP VVV MEYS funded project CZ.02.1.01/0.0/0.0/16\_019/0000765 "Research Center for Informatics". This work has received support from an STFC Grant (ST/T000570/1). The authors would also like to acknowledge the use of the University of Oxford's Advanced Research Computing (ARC) \citet{Richards:2015:ARC} facility in carrying out this work. We acknowledge support of GMRT telescope operators for observations. The GMRT is run by the National Centre for Radio Astrophysics of the Tata Institute of Fundamental Research, India. 


\bibliography{HRMS_arxiv}

\end{document}